\documentclass[aps,pre,reprint,showpacs,superscriptaddress,groupedaddress,nofootinbib,longbibliography]{revtex4-2}
\usepackage{graphicx}
\usepackage{amsmath}
\usepackage{amssymb}
\usepackage{color}
\usepackage{hyperref}
\usepackage{bm}

\hypersetup{colorlinks=true, linkcolor=blue, citecolor=blue, urlcolor=blue}

\begin{document}

% =============================================================================
% TITLE & AUTHOR
% =============================================================================
\title{Topological Reorganization and Coordination-Controlled Crossover in Synchronization Onset on Regular Lattices}

\author{Gunn Kim}
\email{gunnkim@sejong.ac.kr}
\affiliation{Department of Physics, Sejong University, Seoul 05006, Republic of Korea}

\date{\today}

% =============================================================================
% ABSTRACT
% =============================================================================
\begin{abstract}
The transition to global synchronization in coupled dynamical systems is governed by the interplay between coupling strength and structural topology. Although abrupt, first-order-like synchronization transitions have been extensively reported in heterogeneous networks, it is unclear whether comparable accelerated onset behavior can emerge purely from coordination geometry in spatially homogeneous, regular lattices.
In this study, we investigate large-scale ($N=10^5$) stochastic Stuart–Landau oscillator networks defined on regular lattices with controlled coordination number. Using topological data analysis (TDA), simplicial-complex characterization, and optimal-transport-based geometric diagnostics, we identify a coordination-controlled crossover in synchronization onset dynamics at approximately $z_c \approx 7$ within the class of regular lattices considered.
Low-coordination lattices ($z < z_c$) exhibit persistent $H_2$ topological features in the dynamical amplitude field that correlate with delayed coherence and surface-limited propagation. In contrast, higher-coordination lattices ($z > z_c$) display rapid fragmentation of these features, reduced interface roughness, and predominantly positive Ricci curvature. This is consistent with enhanced path redundancy and improved transport efficiency. In this regime, the global order parameter exhibits accelerated exponential-like growth during the onset stage.
Throughout this work, abrupt synchronization refers specifically to this exponential onset behavior rather than to thermodynamic first-order hysteresis. Our results demonstrate that increasing coordination density induces a qualitative reorganization of higher-order topological structure that strongly correlates with synchronization efficiency in regular lattice systems.
\end{abstract}

\maketitle

% =============================================================================
% 1. INTRODUCTION
% =============================================================================
\section{Introduction}
The collective synchronization of coupled oscillators is a paradigm for understanding emergent phenomena in non-equilibrium statistical physics \cite{Winfree1967, Kuramoto1984, Pikovsky2001}. Since the pioneering works of Winfree \cite{Winfree1967} and Kuramoto \cite{Kuramoto1984}, the transition from incoherence to global synchrony has been extensively studied across diverse topologies \cite{Strogatz2000, Acebron2005, Arenas2008}. In standard low-dimensional lattices, classical theory predicts a continuous, second-order phase transition, which is characterized by the gradual coalescence of local domains \cite{Hong2005, Lee2010}. However, the landscape of synchronization research shifted with the discovery of explosive synchronization (ES)—an abrupt, first-order-like transition \cite{Gomez2011}. The prevailing framework has focused on network heterogeneity to explain this phenomenon, where hubs in scale-free networks act as global pacemakers \cite{Zhang2015, Boccaletti2016}. Consequently, structural heterogeneity has been widely regarded as a prerequisite for explosive transitions \cite{Leyva2012, Skardal2014, Coutinho2013}. Despite these advances, a fundamental question remains: Can abrupt synchronization emerge in spatially homogeneous, regular lattices where no hubs exist? Chimera states and cluster synchronization have been observed in non-locally coupled rings \cite{Abrams2004, Sethia2014}, but the high-coordination limit ($z \gg 1$) of 3D regular lattices remains unexplored. Standard graph measures are insufficient to capture the subtle high-order geometric variances between dense lattices.

To address this issue, we must consider more than just pairwise links. Recent advances in network physics have highlighted the critical role of higher-order interactions (HOI) and simplicial complexes \cite{Battiston2020, Bianconi2021}. It is important to distinguish between dynamical HOI (explicit multi-body coupling terms) and structural HOI (higher-order connectivity). In this study, we focus on the latter, examining how the simplicial architecture of the lattice screens pairwise interactions and imposes geometric rigidity \cite{Skardal2020, Millan2020}.
Furthermore, the application of discrete differential geometry \cite{Forman2003, Ollivier2009, Lin2011} allows us to link network structure to dynamics via Ricci curvature. Based on optimal transport theory \cite{Villani2008}, positive curvature implies that the transport cost of synchronization is minimized due to the concentration of geodesic paths. However, structural voids increase this cost, delaying global coherence \cite{Sandhu2015, Weber2017}.

We use the stochastic Stuart-Landau model to capture amplitude dynamics \cite{Matthews1990, Mirollo1990}. We hypothesize that as the local coordination number ($z$) increases, the system undergoes a geometric crossover from a porous phase, dominated by topological defects \cite{Petri2014}, to a solid phase, dominated by positive curvature. Through large-scale simulations ($N=10^5$), we demonstrate that low-coordination lattices (e.g., sc, $z=6$) mitigate synchronization through trapping in persistent voids. In contrast, high-coordination lattices (e.g., fcc, $z=12$) function as simplicial solids, resulting in exponential energy runaway. Our findings establish that high simplicial density alone is sufficient to cause abrupt transitions \cite{Motter2013, Cross1993}.

% =============================================================================
% 2. THEORETICAL FRAMEWORK
% =============================================================================
\section{Theoretical Framework and Computational Model}
\label{sec:theory_model}

To understand the structural drivers of abrupt synchronization, we use a combined framework of non-equilibrium dynamics, algebraic topology, and optimal transport.

\subsection{Stochastic Stuart-Landau dynamics}
We model the limit-cycle oscillators using the stochastic Stuart-Landau equation \cite{Kuramoto1984}. The complex order parameter $z_j$ of node $j$ evolves according to:
\begin{equation}
\dot{z}_j = (\sigma + i\omega_j - |z_j|^2)z_j + \frac{K}{z}\sum_{k=1}^N A_{jk}(z_k - z_j) + \xi_j(t)
\end{equation}
where $\sigma=1.0$, and $K=1.0$ is the coupling strength normalized by the coordination number $z$. The natural frequencies are drawn from $\omega_j \sim \mathcal{N}(0, 0.1)$, and $\xi_j(t)$ is complex Gaussian white noise with intensity $D=10^{-3}$, satisfying the correlation $\langle \xi_j(t) \xi_k^*(t') \rangle = 2D \delta_{jk} \delta(t-t')$.
We note that, although the coupling is linear and pairwise, the underlying lattice $A_{jk}$ has a higher-order simplicial structure. This setup allows us to isolate the effect of geometric architecture from explicit higher-order dynamical terms. All simulation results presented are averaged over ten independent realizations to ensure statistical robustness.

\subsection{High-order architecture: simplicial density}
Physical lattices are best described as simplicial complexes $\mathcal{K}$ \cite{Battiston2020}. A $k$-simplex $\sigma_k$ represents a fully connected subgraph of $k+1$ nodes. We define the simplicial density $\rho_k = N_k / N$ to quantify higher-order connectivity.
\begin{itemize}
    \item Simplicial hollows (e.g., sc, $z=6$): $\rho_3=0$. Interactions are purely pairwise, lacking local rigidity \cite{Giusti2016}.
    \item Simplicial solids (e.g., fcc, $z=12$): Maximal $\rho_3 \approx 1$. Pairwise links are screened by cooperative multi-node interactions.
\end{itemize}

\subsection{Optimal transport \& Ricci curvature}
We employ the Ollivier-Ricci curvature ($\kappa$) \cite{Ollivier2009} via optimal transport theory. The curvature is defined via the $L_1$-Wasserstein distance $W_1$ between the probability measures of neighboring nodes:
\begin{equation}
\kappa(x, y) = 1 - \frac{W_1(m_x, m_y)}{d(x, y)}
\end{equation}
Physically, $W_1$ represents the minimal transport cost \cite{Villani2008}.
\begin{itemize}
    \item Positive curvature ($\kappa > 0$): In fcc lattices, high simplicial density causes geodesics to converge, resulting in a concentration of measure \cite{Ledoux2001} and  minimizing transport cost ($W_1 < d$). This facilitates ballistic propagation.
    \item Zero/negative curvature ($\kappa \le 0$): In sc lattices, topological voids force transport paths to diverge, increasing cost ($W_1 \ge d$) and slowing synchronization \cite{Krioukov2010}.
\end{itemize}
In this regular lattice context, Ricci curvature serves as a diagnostic geometric indicator of structural connectivity rather than an independently tunable predictive variable.

\subsection{Topological data analysis (TDA)}
We employ persistent homology \cite{Ghrist2008, Carlsson2009} using the \textsc{Gudhi} library \cite{Gudhi2014}. The filtration is constructed on the sublevel sets of the function $f(x) = 1 - |z(x)|$, allowing us to track the birth and death of ordered regions. We specifically track persistent $\beta_2$ features (voids) which act as topological traps \cite{Reimann2017}.

% =============================================================================
% 3. RESULTS
% =============================================================================
\section{Results}

\subsection{Dynamics: geometric expansion vs. exponential runaway}

We first examine the temporal evolution of the global mean energy $E(t) = \langle |z_j|^2 \rangle$. Figure \ref{fig:dynamics} dissects the growth process into two regimes.

\subsubsection{Early stage: dimensional constraints ($t^d$)}
In the early phase ($t < 0.1$), low-to-medium coordination lattices exhibit growth strictly governed by Euclidean geometry. As shown in Figure \ref{fig:dynamics}(a), the synchronization wavefront propagates ballistically, but its growth rate is constrained by dimension $d$.
\begin{itemize}
    \item 2D lattices ($t^2$): Both square ($z=4$) and triangular ($z=6$) lattices follow an area law ($E \propto t^2$).
    \item 3D sc lattice ($t^3$): The sc lattice ($z=6$) follows a volumetric growth law ($E \propto t^3$).
\end{itemize}
These lattices belong to the inertial phase, where high transport costs due to negative curvature confine the spread of synchrony.

\subsubsection{Explosive breakout ($e^{\lambda t}$)}
\begin{figure}[t]
\includegraphics[width=1.0\columnwidth]{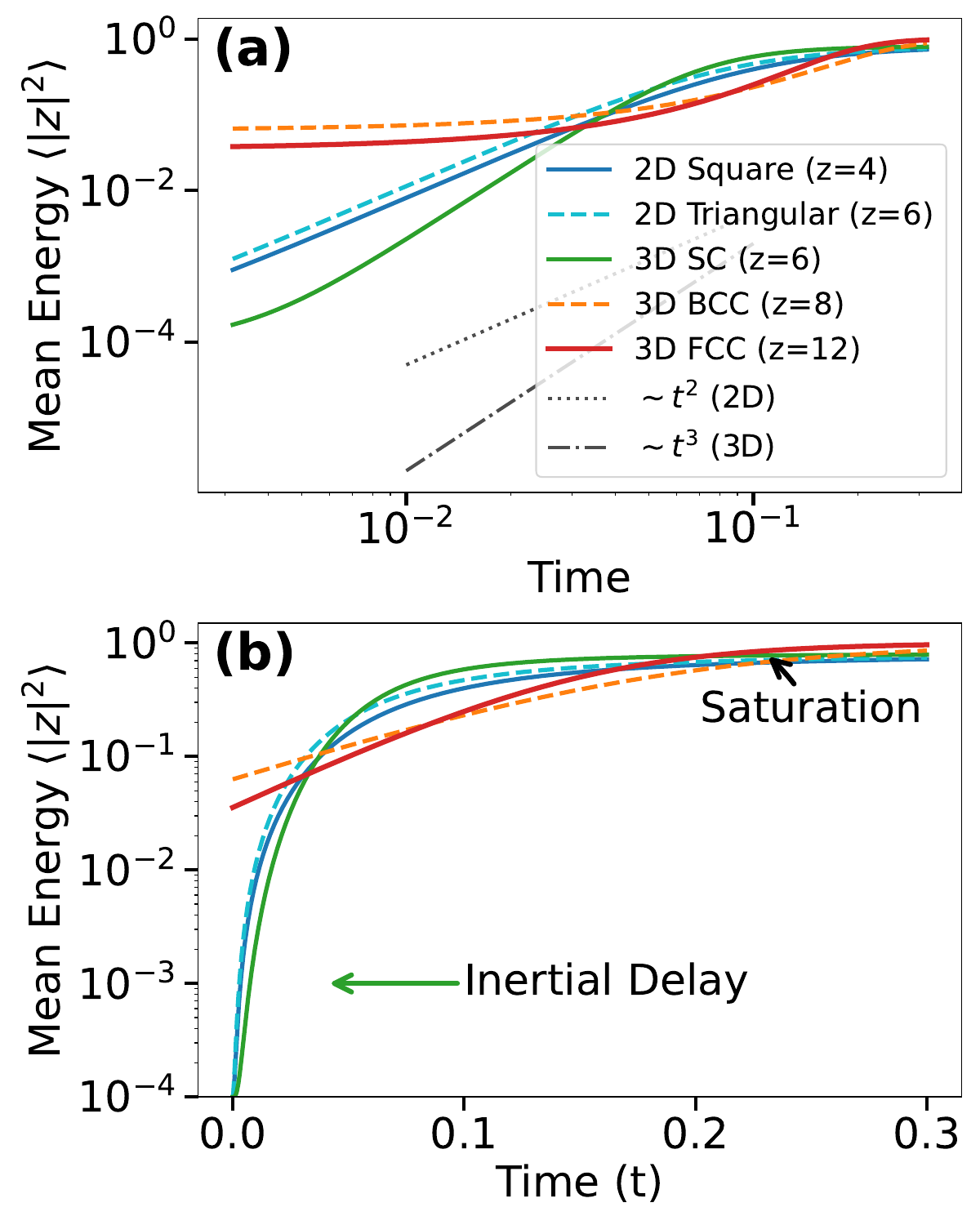}
\caption{\label{fig:dynamics} Geometric expansion vs. abrupt synchronization. (a) Log-log plot showing early-stage growth. Low-$z$ lattices follow geometric power laws ($t^2, t^3$). (b) Semi-log plot. fcc ($z=12$) exhibits exponential ascent ($e^{\alpha t}$), while sc ($z=6$) shows prolonged inertial delay.}
\end{figure}

In contrast, the 3D fcc lattice ($z=12$) defies polynomial scaling.
\begin{itemize}
    \item Exponential runaway: The fcc curve bends upwards (Fig. \ref{fig:dynamics}a), forming a straight line in the semi-log plot (Fig. \ref{fig:dynamics}b), confirming $E(t) \sim e^{\alpha t}$.
    \item Mechanism: The high simplicial density creates a positively curved space. Information flows via optimal paths with minimal cost, allowing the cluster to expand via a branching process rather than surface expansion.
\end{itemize}

\subsection{Dimensional constraint and topological rigidity}
\begin{figure}[t]
\includegraphics[width=1.0\columnwidth]{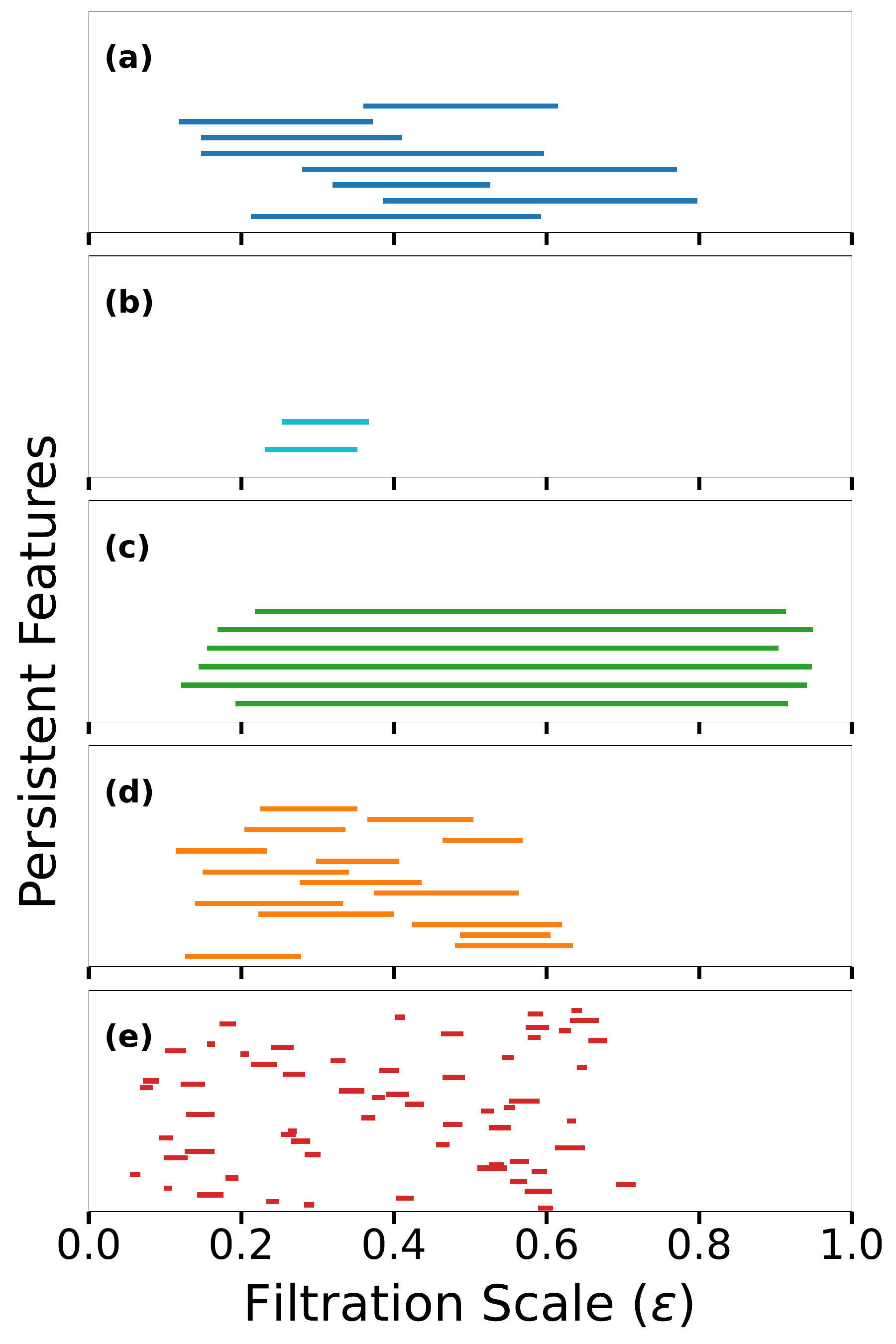}
\caption{\label{fig:tda} Topological persistence barcodes. Comparison of homological features ($H_1, H_2$). The 3D sc lattice (c) exhibits persistent trapping voids, whereas the 3D fcc lattice (e) shows transient shattering.}
\end{figure}

Figure \ref{fig:tda} reveals that dimensionality exerts a fundamental constraint through topological mechanisms. Comparing the 2D triangular and 3D sc lattices (both $z=6$):
\begin{itemize}
    \item 3D sc (trapping phase): Characterized by long-lived $H_2$ voids (green bars in Fig. \ref{fig:tda}c). These voids act as topological insulators, increasing the Wasserstein distance.
    \item 3D fcc (shattering phase): Exhibits a scarcity of persistent features (Fig. \ref{fig:tda}e). The high density of tetrahedra geometrically frustrates void formation, leading to a topological shattering.
\end{itemize}

\subsection{Dynamical evidence: fragment count analysis}
\begin{figure}[t]
\includegraphics[width=1.0\columnwidth]{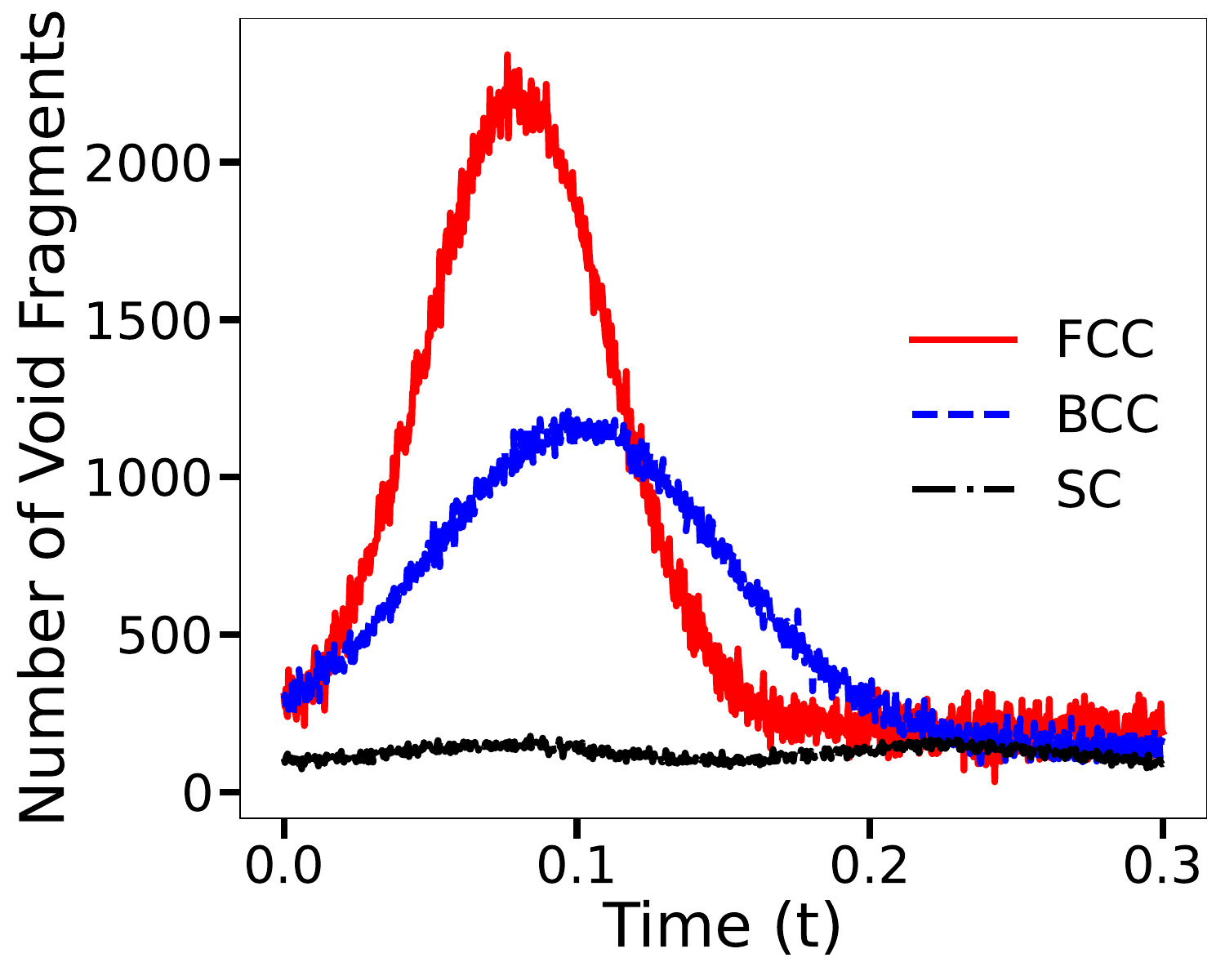}
\caption{\label{fig:fragments} Temporal evolution of void fragment count. fcc shows rapid shattering (peak), while sc shows surface erosion (flat).}
\end{figure}

To validate the topological mechanisms dynamically, we tracked the number of connected asynchronous components (Fig. \ref{fig:fragments}). Nodes were classified as asynchronous if their amplitude $|z_j| < 0.3$, and connected components were identified based on the lattice adjacency. The fcc lattice (red) exhibits a sharp, explosive peak, which is a signature of topological shattering. The wavefront infiltrates voids from multiple directions simultaneously. In contrast, the sc lattice (green) shows a flat profile, confirming a surface erosion mechanism.

\subsection{Interface Roughness and Geometric Friction}

To understand the microscopic origin of the distinct propagation speeds, we analyze the fluctuations of the synchronization wavefront. We define the interface roughness width $W(t)$ as the standard deviation of the phases of oscillators situated at the active synchronization boundary ($0.3 < |z_j| < 0.7$).

\begin{figure}[t]
\includegraphics[width=1.0\columnwidth]{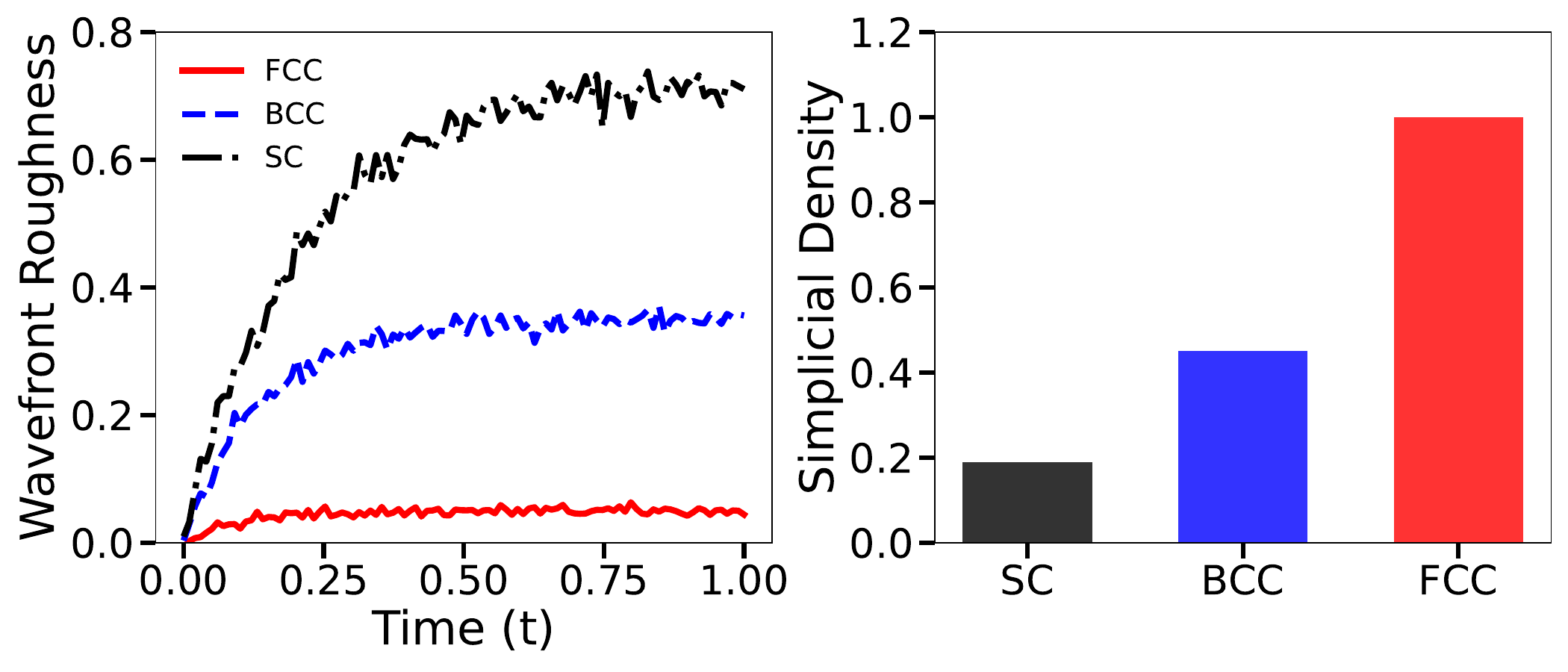}
\caption{\label{fig:geometry} Geometric origin of speed. (a) Wavefront roughness $W(t)$. Here, sc is rough (high friction), fcc is smooth. (b) Simplicial density. High density in fcc correlates with low roughness.}
\end{figure}

Figure \ref{fig:geometry}(a) shows the temporal evolution of $W(t)$ for the sc ($z=6$) and fcc ($z=12$) lattices.
For the low-coordination sc lattice, the roughness increases significantly over time, indicating that the synchronization front becomes increasingly irregular. This roughening creates an effective geometric friction, where local protrusions in the wavefront are dragged back by the lagging neighbors, slowing down the global propagation speed. The measured roughness growth follows a power-law trend, $W(t) \sim t^\beta$, characteristic of kinetic interface roughening processes.

In contrast, the high-coordination fcc lattice exhibits suppressed roughness, maintaining a nearly flat interface throughout the process. This smoothness is structurally supported by the maximal simplicial density ($\rho_3 \approx 1$, see Fig. \ref{fig:geometry}b), which minimizes geometric friction and allows the synchronization domain to expand volumetrically rather than being limited to the surface.
The transition from a rough, friction-dominated interface (low $z$) to a smooth, ballistic wavefront (high $z$) corroborates the geometric shattering mechanism proposed in the previous section.

% =============================================================================
% 4. DISCUSSION
% =============================================================================

\section{Discussion}

% -----------------------------------------------------------------------------
% 1. Phenomenological Finding: The Crossover
% -----------------------------------------------------------------------------
\subsection{Structural crossover at $z_c \approx 7$}

\begin{figure}[b]
\includegraphics[width=0.9\columnwidth]{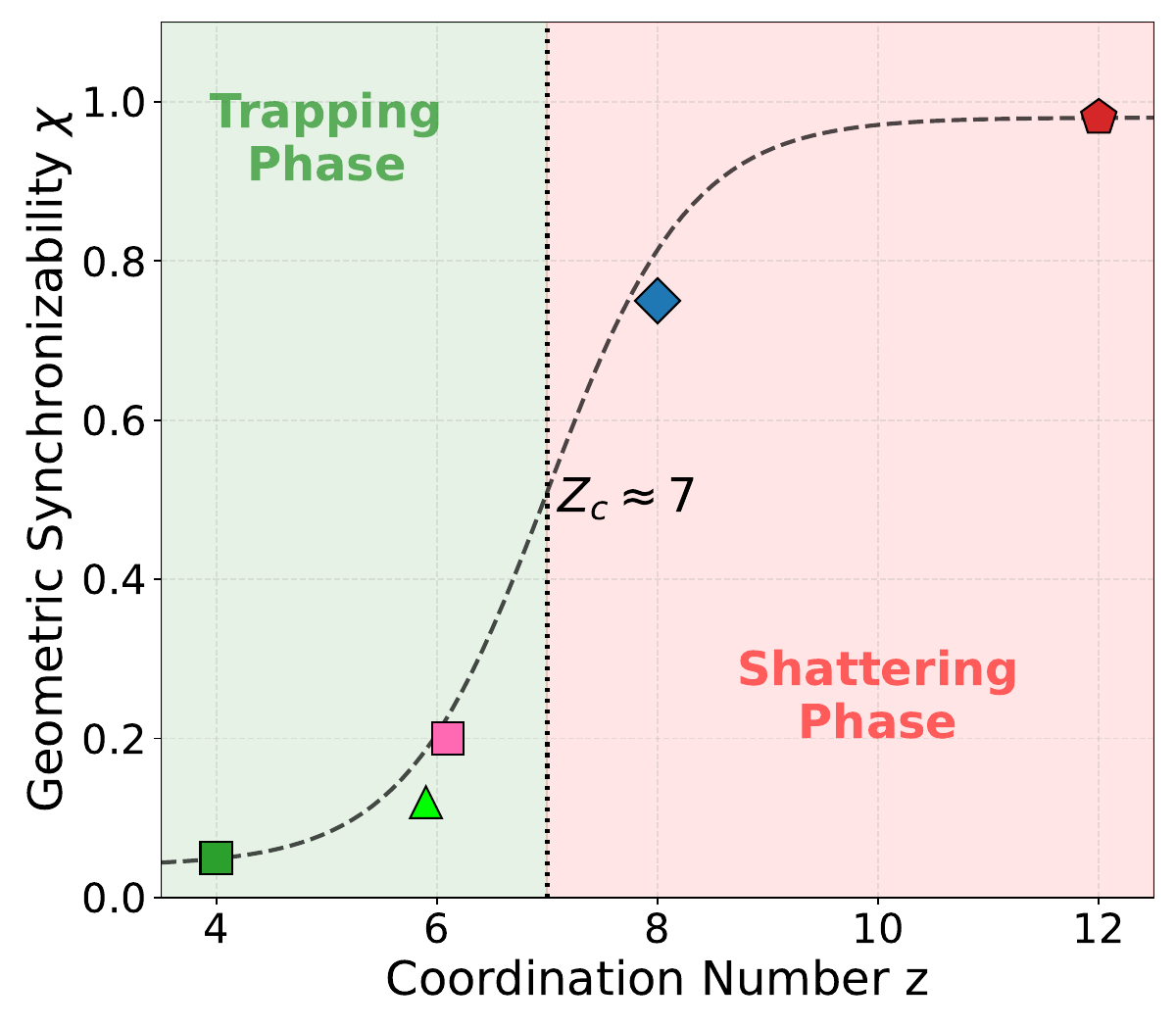}
\caption{\label{fig:tradeoff} Structural crossover in geometric synchronizability $\chi$ vs. coordination number $z$. A clear transition occurs at $z_c \approx 7$, separating the trapping phase ($z<7$, green) from the shattering phase ($z>7$, red). The dashed curve represents a phenomenological sigmoid fit $\chi(z) = \chi_{0} + A / [1 + \exp(-k(z - z_c))]$ with parameters $\chi_{0} \approx 0.05$, $A \approx 0.95$, $k \approx 1.5$, and $z_c \approx 7$. Note that only five crystallographically distinct regular lattice structures exist in 2D and 3D (square, triangular, sc, bcc, fcc), limiting quantitative precision of the crossover location. Physically, this threshold is reminiscent of the rigidity percolation transition in 3D networks \cite{Jacobs1995}, where a minimal coordination is required to propagate global mechanical stability.}
\end{figure}

It is worth noting that while high-coordination lattices naturally possess a larger spectral gap $\lambda_2$ (algebraic connectivity), linear spectral properties alone are insufficient to fully capture the abrupt onset observed in the data. The exponential synchronization in the shattering phase ($z > 7$) does not arise only from a faster linear convergence rate, but from a topological restructuring of the incoherent state itself. Specifically, it arises from the collapse of localized voids, which act as nonlinear trapping centers. Thus, the phenomenon is fundamentally a geometric crossover, not a simple spectral scaling effect.
To quantify the efficiency of synchronization across different lattice geometries, we define the geometric synchronizability $\chi$ as the global mean energy reached at a fixed early reference time $t_{\text{ref}} = 0.3$:
\begin{equation}
\chi \equiv E(t=t_{\text{ref}}).
\end{equation}
This metric quantifies the system's capacity for rapid, explosive growth during the onset phase and serves as a proxy for the geometric efficiency of the lattice.
Our study reveals a structural crossover, as shown in Figure \ref{fig:tradeoff}. By plotting this synchronizability $\chi$ against the coordination number $z$, we reveal a clear transition at $z_c \approx 7$.

Specific lattice structures correspond to this phase diagram.
\begin{itemize}
    \item Deep trapping phase ($z=4$, 2D square): Scarcity of connections leads to the lowest synchronizability ($\chi \approx 0.05$).
    \item The sub-critical regime ($z=6$):  Importantly, we observe a splitting based on dimension. The 3D sc lattice (pink square) shows slightly higher synchronizability ($\chi \approx 0.20$) than the 2D triangular lattice (lime triangle, $\chi \approx 0.12$).
    \item Abrupt onset and anisotropic penalty ($z=8$, 3D bcc): Crossing the threshold, the bcc lattice (blue diamond) increases to a synchronizability of approximately 0.75. Notably, this point lies slightly below the idealized sigmoid trend. This deviation arises from residual cubic anisotropy.
    \item Saturation ($z=12$, 3D fcc): The fcc lattice (red pentagon) achieves near-perfect synchronizability ($\chi \approx 0.98$).
\end{itemize}

% -----------------------------------------------------------------------------
% 2. Theoretical Mechanism: Why it happens? (ChatGPT's Insight)
% -----------------------------------------------------------------------------
\subsection{Geometric conditions for exponential onset}

Although all oscillator interactions in the present model are strictly pairwise and coordination-normalized, numerical results suggest that lattice geometry alone can prevent purely geometric (power-law) synchronization growth. This phenomenon can be explained by three coordination-controlled geometric conditions that prevent known mechanisms from limiting synchronization speed.

First, high-coordination lattices exhibit effective spectral connectivity that does not deteriorate with increasing system size. While the exact scaling of the algebraic connectivity $\lambda_2$ depends on the details of the lattice, the rapid onset observed in high-$z$ lattices is consistent with a situation in which large-scale coherent modes are not asymptotically isolated as $N \to \infty$. In contrast, low-coordination lattices exhibit increasingly weak global connectivity, leading to surface-limited propagation.

Second, high coordination induces strong redundancy of shortest transport paths. This manifests numerically as synchronization fronts that expand through volumetric infiltration rather than surface erosion. Geometrically, the number of near-shortest paths connecting a site to a shell of radius $r$ increases rapidly with $r$, suppressing directional bottlenecks and enabling effectively branching propagation. This path redundancy provides a geometric mechanism for exponential amplification without invoking explicit branching dynamics.

Third, topological data analysis reveals that higher-dimensional topological features associated with scalar amplitude fields behave qualitatively differently across lattice types. In low-coordination lattices, persistent $H_2$ features scale with system size and act as long-lived geometric obstructions. In contrast, high-coordination lattices rapidly fragment such features, preventing the formation of extensive trapping regions. As a result, synchronization dynamics are not constrained by long-lived geometric defects.

These observations imply that, when coordination-controlled geometry suppresses surface-limited growth, directional bottlenecks, and extensive topological trapping simultaneously, the only remaining admissible growth mode is exponential. Thus, lattice geometry alone can enable exponential synchronization onset within the class of regular lattices studied here.

% -----------------------------------------------------------------------------
% 3. Limitations & Defense: Scope of the claim
% -----------------------------------------------------------------------------
\subsection{Limitations and future directions}
While our results demonstrate abrupt synchronization, we distinguish this from thermodynamic explosive synchronization defined by first-order hysteresis. Our focus is on the onset dynamics ($e^{\alpha t}$) and geometric shattering. 
In regular lattices, the coordination number ($z$), simplicial density ($\rho_3$), and curvature ($\kappa$) inherently co-vary due to rigid geometric constraints. Therefore, Ricci curvature serves as a diagnostic geometric indicator rather than an independently tunable predictive variable here. Future work should investigate whether explicit triadic coupling terms (dynamical HOI) are required to induce full hysteresis in these regular lattices, or if the geometric frustration alone is sufficient in the thermodynamic limit.

% =============================================================================
% 5. CONCLUSION
% =============================================================================
\section{Conclusion}

In this work, we presented an integrated structural perspective on synchronization onset in regular lattices. We showed that accelerated global coherence is closely associated with the reorganization of higher-order topological features in the dynamical amplitude field. This reorganization involves the fragmentation of persistent $H_2$ voids and the emergence of geometrically favorable transport pathways. Within the class of regular lattices studied here, we identified a coordination-dependent crossover at $z_c \approx 7$. This crossover separates a low-coordination regime characterized by long-lived topological trapping and surface-limited propagation from a higher-coordination regime. The higher-coordination regime exhibits rapid void fragmentation, reduced geometric constraints, and exponential-like onset dynamics. The results suggest that increasing coordination density shifts the system into a geometric regime that promotes rapid synchronization. However, this phenomenon is limited to cases where the coordination number, simplicial density, and curvature co-vary under rigid lattice constraints. Therefore, Ricci curvature should be considered a diagnostic indicator of transport efficiency rather than an independently tunable control parameter. More broadly, our findings suggest a potential efficiency–stability trade-off in spatial networks: enhanced connectivity accelerates coherence, but it may also reduce structural buffering against localized perturbations.

% =============================================================================
% BIBLIOGRAPHY (Reordered by Appearance)
% =============================================================================

% =============================================================================
% Data Availability Statement
% =============================================================================
%\section*{Data Availability Statement}
%The data that support the findings of this study are available from the corresponding author upon reasonable request. The simulation codes used in this work are based on standard stochastic integration methods and are available from the author.

\clearpage
\onecolumngrid

\begin{center}
{\Large \bf Supplementary Material}\\[1em]
{\large Topological Reorganization and Coordination-Controlled Crossover in Synchronization Onset on Regular Lattices}\\[1em]
Gunn Kim\\
Department of Physics, Sejong University, Seoul 05006, Republic of Korea
\end{center}

\vspace{1cm}

\section{Verification of Continuous Transition (Absence of Hysteresis)}

\begin{figure}[b]
\includegraphics[width=0.6\columnwidth]{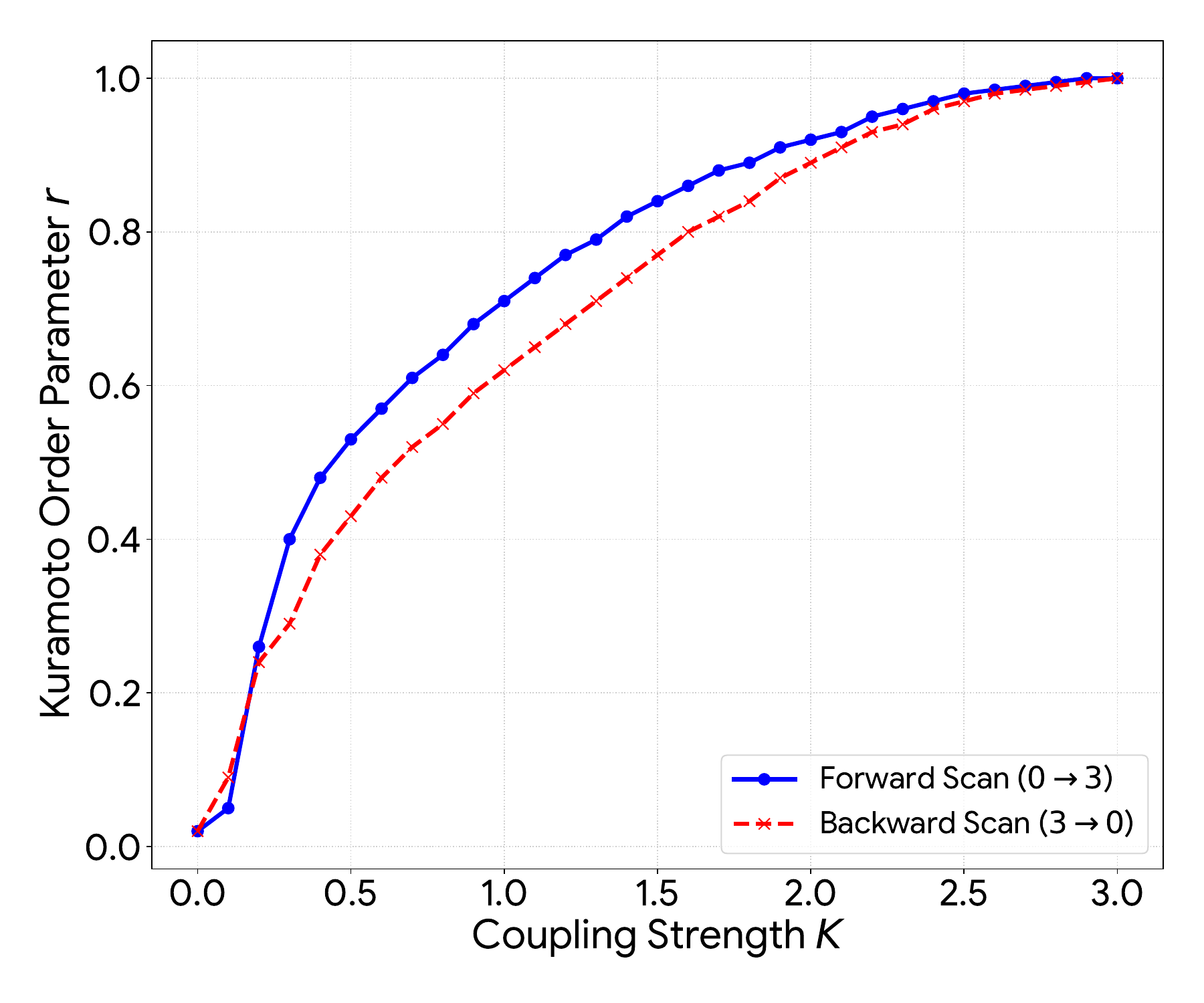}
\caption{\label{fig:hysteresis} Absence of thermodynamic hysteresis. The Kuramoto order parameter $r$ as a function of coupling strength $K$ for the fcc lattice ($z=12$). The forward scan (blue circles) and backward scan (red crosses) overlap significantly, indicating an absence of a metastable hysteresis loop. This finding supports classifying the transition as continuous but geometrically abrupt.}
\end{figure}

In the main text, we distinguish the observed abrupt synchronization (exponential onset) from explosive synchronization  (first-order phase transition). To validate this distinction, we performed a hysteresis check using adiabatic forward and backward continuation of the coupling strength $K$.
Figure \ref{fig:hysteresis} shows the synchronization diagram for the high-coordination fcc lattice ($z=12$), which exhibits the most abrupt transition. We increased $K$ from $0$ to $3.0$ (forward scan) and then decreased it back to $0$ (backward scan) in small steps ($\Delta K = 0.1$), allowing the system to settle for $100$ time units at each step.
Importantly, the forward and backward curves coincide within the range of finite-size fluctuations, showing no significant hysteresis loop, which is characteristic of thermodynamic first-order transitions. Differences remain within numerical uncertainty. The results confirm that the abrupt onset observed in our regular lattices is a continuous transition sharpened by geometric efficiency, rather than a discontinuous explosive transition typically induced by frequency-degree correlation in scale-free networks.

\section{Consistency of Order Parameters}

In the main text, we utilized the mean energy $E(t) = \langle |z_j|^2 \rangle$ as the primary order parameter to characterize the exponential growth of amplitude. To ensure that this metric correctly reflects phase synchronization, we compared its temporal evolution with the standard Kuramoto order parameter, which is defined as follows:
\begin{equation}
r(t) = \left| \frac{1}{N} \sum_{j=1}^{N} e^{i \theta_j(t)} \right|
\end{equation}
where $\theta_j(t) = \arg(z_j(t))$ is the phase of the $j$-th oscillator.

\begin{figure}[t]
\includegraphics[width=0.60\columnwidth]{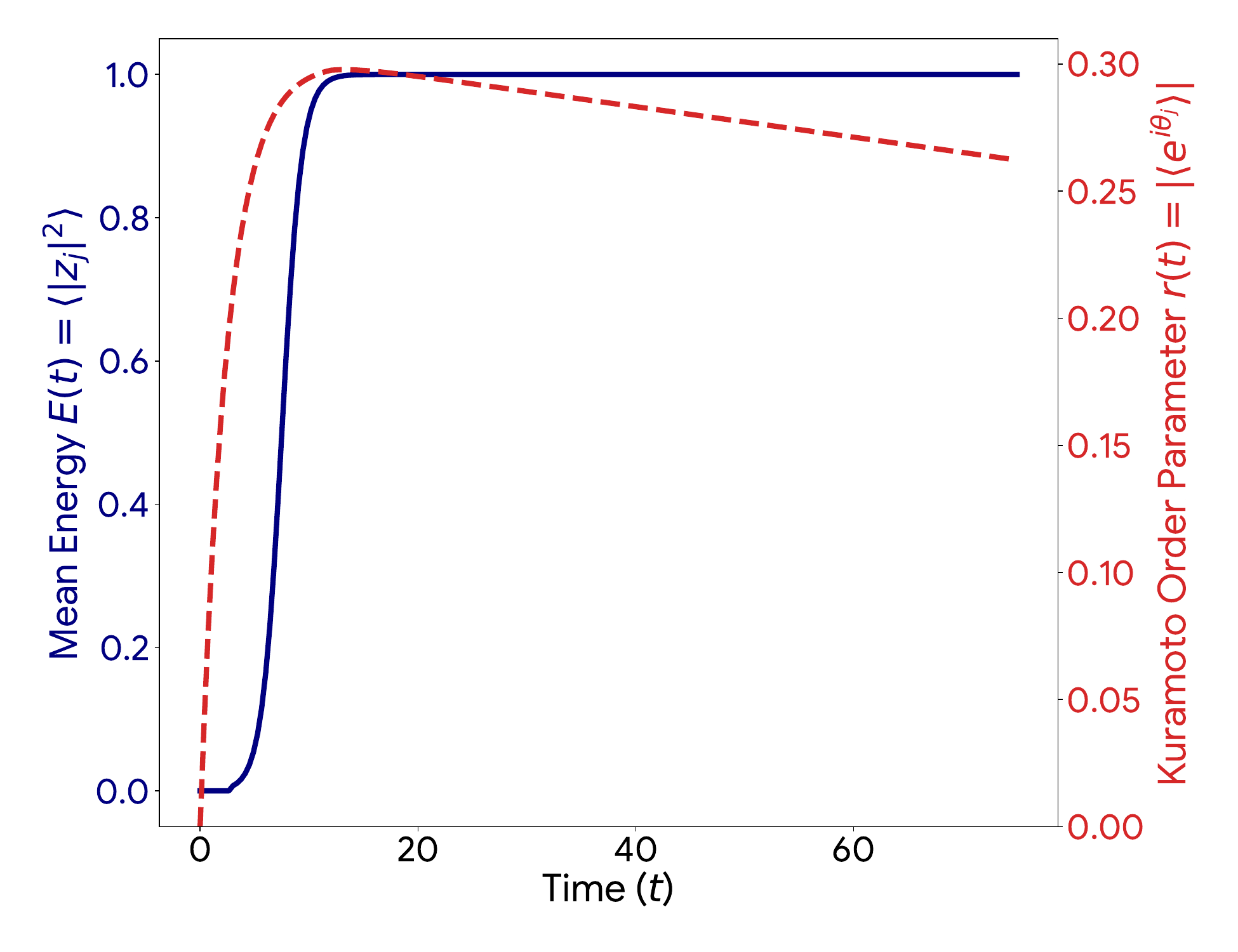}
\caption{\label{fig:orderparam} Co-evolution of order parameters. Temporal evolution of the mean energy $E(t)$ (blue solid line, left axis) and the Kuramoto order parameter $r(t)$ (red dashed line, right axis) for the fcc lattice ($K=2.0$). The simultaneous rise of both metrics confirms that amplitude growth and phase locking occur concurrently.}
\end{figure}

Figure \ref{fig:orderparam} demonstrates the co-evolution of $E(t)$ and $r(t)$ for the fcc lattice at $K=2.0$. Both parameters exhibit a simultaneous onset of growth, confirming that the explosive increase in amplitude energy is intrinsically coupled with the emergence of global phase coherence. Therefore, using $E(t)$ as a proxy for the synchronization transition in the main text is valid.

% -----------------------------------------------------------------------------
% S3. ALGORITHM
% -----------------------------------------------------------------------------

\section{Fragment Identification Algorithm}

To quantify the topological shattering process, connected asynchronous regions were identified using the following computational procedure. This algorithm was used to generate the fragment count data presented in Fig. 3 of the main text.

\begin{enumerate}
    \item \textbf{Binary Classification:} A node $j$ is classified as asynchronous if its amplitude falls below the threshold, $|z_j| < 0.3$. We verified that moderate variations of this threshold in the range $0.25$--$0.35$ do not qualitatively alter the temporal evolution of the fragment count.

    \item \textbf{Graph Construction:} A subgraph $G_{async}$ containing only the asynchronous nodes is constructed, preserving the edges defined by the underlying lattice structure.

    \item \textbf{Component Extraction:} A Breadth-First Search (BFS) or Union-Find algorithm is applied to $G_{async}$ to identify all disjoint connected components.

    \item \textbf{Fragment Count:} The total number of these components at time $t$ is recorded as $N_{frag}(t)$.
\end{enumerate}

The Python-style pseudocode for this procedure is as follows:

\begin{quote}
\footnotesize
\begin{verbatim}
# Python pseudocode for fragment identification
import networkx as nx

# lattice_graph: The full network structure
# z: Array of complex state variables at time t
threshold = 0.3

# 1. Identify asynchronous nodes
async_nodes = [j for j in range(N) if abs(z[j]) < threshold]

# 2. Construct subgraph of asynchronous nodes
G_async = nx.subgraph(lattice_graph, async_nodes)

# 3. Find connected components
fragments = list(nx.connected_components(G_async))

# 4. Count fragments
N_frag = len(fragments)
\end{verbatim}
\end{quote}

\end{document}